\documentclass [Coference]{IEEEtran}
\usepackage[dvips]{graphicx}
\usepackage{tikz}
\usepackage{amsmath,amssymb,amsfonts,amsthm}
\usepackage{colortbl,longtable,subfig}
\usepackage{color}
\newtheorem{thm}{Theorem}

\newtheorem{lem}{Lemma}

\usepackage[normalem]{ulem}
\usepackage{soul}
\usepackage{setspace}

\usepackage{bm}

\usepackage{framed}

\usepackage{ifthen}

\newcommand{\TR}{\ifthenelse{1<0}}

\newcommand{\eop}{\hfill{$\blacksquare$} }

\newcommand{\ignore}[1]{}

\newcommand{\beq}{\begin{eqnarray*}}
\newcommand{\eeq}{
\end{eqnarray*}}

\newcommand{\bnd}{\beta}

\begin{document}
 \title{Acquisition Games with Partial-Asymmetric  Information }
 \author{$^1$Veeraruna Kavitha,  $^1$Mayank Maheshwari and $^2$Eitan Altman \\
 $^1$IEOR, Indian Institute of Technology Bombay, India  and  $^2$INRIA, France}
 \maketitle

 \renewcommand{\a }{{a}}
 \newcommand{\dtau}{\delta}

\newcommand{\btau}{ {\bar \tau}}
\newcommand{\ba}{ {\bar a}}
\renewcommand{\a}{{a}} 
\newcommand{\x}{{x}}

\newcommand{\z}{{\bf z}} 
\renewcommand{\t}{{\bm \tau}}

\begin{abstract}

\footnote{The work is partially sponsored by MALENA, the joint research team between IIT Bombay and Inria.}We consider an example of  stochastic games with partial,  asymmetric and non-classical information. We obtain relevant equilibrium policies using a new approach which allows managing the belief updates in a structured manner.  Agents have access only to partial information updates, and our approach is to consider optimal open loop control until the information update. 
The agents continuously control the rates of their Poisson search clocks to acquire the locks, the agent to get all the locks before others would get reward one.   However, the agents have no information about the acquisition status of others and will incur a cost proportional to their rate process. We solved the problem for the case with two agents and two locks and conjectured the results for $N$-agents. 
  We showed that a pair of  (partial) state-dependent time-threshold policies form a Nash equilibrium. 

\end{abstract}

\section{Introduction}
We consider  non-classical and asymmetric information (as specified in \cite{basar}) based games inspired by the full information   games  considered in \cite{eitan}.  
In \cite{eitan},   agents attempt to acquire     $M$ available destinations;  each agent controls its rate of advertisement through a Social network  to increase its chances of winning the destinations, while trading off the cost for acquisition.   They considered full information and no-information games, and considered discrete time policies by uniformization of the controlled Markov process.  In full information games,  the agents at any point of time  know the number of available destinations or equivalently the number of destinations already acquired by one or the other agent. In no-information games, the agents have no information; they don't even know the number of destinations acquired by themselves. 
 
 It is more realistic to assume that the agents know the number of destinations acquired by themselves, but would not know the number acquired by  others.  This leads to partial, asymmetric and non-classical information games, which are  the main focus of this paper.  Basar et. al in \cite{basar} describe a game to be of non-classical information type, and we describe the same in our   words:  if  
 the state of  agent $i$  depends upon the actions of   agent $j$, and if agent $j$ knows  some information which is not available to agent $i$ we have a non-classical information game.
 These kind of games are typically hard to solve (\cite{basar}); when one attempts to find best response against a strategy profile of others, one would require belief of  others states, belief about the belief of others, and so on.

Our approach to this problem can be summarized as ``open loop control till information update".  With no-information, one has to resort to open loop policies. This is the best when one has no access to information updates.  With full information one can have closed loop policies,  where the optimal action depends upon the state of the system at the decision epoch. In full information    controlled  Markov jump processes, every agent is informed immediately of the jump in the state and can change its action based on the change.  In our case we have access to partial information,  the agents can observe only some jumps and not all; 
thus
we need policies that are open loop type till  an information update.   At every information update, one can choose a new open loop control depending upon the new information.

We considered one and two  lock acquisition problems, any agent wins reward one if it acquires all the locks and if it is the first one to acquire the locks.  The agents have no access to the information/state  of the others,  however upon contacting a lock they would know if they are the first to contact.  We obtained Nash equilibrium for these partial, asymmetric information games; a pair of (partial) state-dependent time threshold policies form Nash equilibrium. We obtained these results (basically best responses) by solving Dynamic programming equations applicable to (partial) information update epochs  and each stage of the Dynamic programming equations are solved by solving appropriate optimal control problems and the corresponding Hamiltonian Jacobi equations. 
 
A block chain network is a distributed ledger that is completely open to  all nodes; each node will have a copy of transactions (in case of currency exchange).    If a new  transaction is  to be added (linked) to the previously existing chain in the form of a new block, it  requires the miners (designated special nodes) to search for a key (encryption), that enables it to be added to the ledger. This search of the key requires computational power and time. The first miner to get the right key,   gets a financial reward.  If  the miners would not   know the status of the search efforts of others, the resulting game is exactly as in our one lock problem. Two lock problem can be viewed as the extension of one lock problem, wherein a second key is required to gain the reward.

\section{Problem Description}

Two agents are  competing  to win a project. There are two  or one  lock(s) to win the project, 
 and the  aim of the agents is to reach these  as quickly as possible. The agent that contacts all the locks before the other   gets a reward equal to one unit.  Further they need to contact the lock(s) before the deadline $T$. 
 
 The contact process is controllable;  the agents control the rate of the contact process continuously over time and they would incur some cost for acceleration.   
       The  acquisition/contact  process is modelled by a Poisson process. 
     The rate of contact process can be time varying over the interval $[0, T]$, it can further depend upon the number of locks acquired etc.  The higher the rate, the higher the chances of success, but also higher is the cost of acquisition. 

{\bf Information structure:} The agents have partial/asymmetric information about  the   locks acquired by various agents and would use   available information to design their acceleration strategies optimally. The agents would know at all the times information (contacted/not contacted etc., at a given time)  related to  its contact attempts, however it has limited access to that of  the others. When any agent contacts a  lock, it would know if it is successful; 
we call a contact successful if the agent is the first one to contact that particular lock.  
If the other agent had contacted the same lock before the tagged customer,  the tagged customer  would have an unsuccessful contact. 
The agent gets an update of this information  immediately after a contact,  and   this  will also reveal some information about the status of the other agents.  For example, upon a contact, if it gets aware  of a successful contact, it also gets to know that this is the first one to contact. 

{\bf Decision epochs:} 
Every agent has a continuous (at all time instances) update of the status of  its contact process, however there is a major update in its information only at (lock) contact instances.  
At these epochs it would know if the contact is successful/unsuccessful which in turn would reveal some information about the state of the other agents.
 Hence these    form the natural decision epochs;   an agent can choose a different action at such time points. Further, it is clear that the  decision epochs of different agents are not synchronous.


{\bf Actions:} The agents should choose an appropriate rate (function) for the contact/acceleration process. 
The rate of contact, for agent $i$,  at any time point can be between $[0, \bnd^i]$.   
The agents choose an action which specifies the acceleration process at the beginning, and change their action  every time  it contacts a lock (successfully or unsuccessfully). 
The action at any decision epoch is a measurable acceleration process (that can potentially span over time interval $[0, T]$). To be precise agent $i$ at decision epoch $k$ (the  instance at which it contacted the $(k-1)$-customer) chooses  an ${a}_{k}^i \in L^\infty [0, T]$, as the acceleration process to be used till the next acquisition. Here  $L^\infty [0, T]$ is the space of all measurable functions that are   essentially bounded, i.e.,  the space of functions with   finite essential supremum norm: 
 $$
|| \a ||_\infty :=  \inf \{ \bnd:  |a(t)| \le \bnd  \mbox{ for allmost all } t \in [0, T]\}.
 $$

{\bf State:} 
We will have two decision epochs with two lock problem, and one decision epoch with one lock problem,  and have corresponding number of major state updates. 
Let ${\z}_k^i$ represent the information available to agent $i$ immediately after $(k-1)$-th contact\footnote{By convention, the start of the process commences with 0-th contact.}.  
Here  ${\z}_k^i$ has two components: a) first component  is a flag indicating that the contact was successful; b) second component is the time of contact.
The first decision epoch (the only decision epoch with one lock problem) is at time 0, and the state $\z_1$ is simply set to  $(0, 0)$  to indicate 0  contacts and     '0' contact time   (which is of no importance since there is no contact yet).  The state at the second decision epoch $\z_2 = (f, \tau)$,  where flag $f = s$  implies successful contact and $f=u$ implies unsuccessful contact while $\tau$ represents the time of  first contact.  Let   $\tau^i_k$ represent the (random)  $k$-th contact instance of agent $i$.  Here we view $\tau_k^i$ as a fictitious random contact instance which can take any value from $[0, \infty)$ and with $\tau_k^i > T- \tau_{k-1}^i$ indicating that the agent could not contact the $k$-th lock before deadline. 
  
 {\it 
We  distinguish the one lock problem from the two lock problem by using $M$ to represent the number of locks. }
 
 {\bf Strategy:} The strategy of player $i$ 
 $$
 \pi^i = \left  \{  \a_{k}^i (\z_k) ; \mbox{ for all possible } \z_k  \right \}_{k  \le M },
 $$where $\a_1 (\cdot)$ represents the acceleration process used at  start, $\a_2 (\cdot)$ represents the acceleration process used after contacting one lock   and this choice depends upon the  available information $\z_2.$  To keep notations simple, most of the times and unless clarity is required, state $\z_k$ is not used while representing the actions. One can easily extend this theory to a more general problem with $M$ locks, but the emphasis of this paper would be on the case with $M=2$.   We     briefly discuss the extension to a larger $M$ towards the end of the paper. 

{\bf Rewards/Costs:} The reward of any agent is one, only if it succeeds to contact  all $M$  locks,  and further,  if it is the first one to contact the first lock.  
Let    $d_1^i = T\wedge \tau^i_1$ ($\wedge$ implies  minimum of the two) and $d_2^i = ( (T\wedge \tau^i_2) - \tau^i_1 )^+$  respectively represent the  durations\footnote{As already mentioned,  the contact clocks $\{\tau^i_k\}$ are free running Poisson clocks, however we would be interested only in those contacts that occurred  before deadline $T$.}  of the first  and the second  contact process. 
The cost spent on acceleration  for the  $k$-th contact equals,  
\begin{eqnarray}
\ba_k^i (d_k^i ), \mbox{ with }   \ba^i_k  (s)   :=  \int_0^s a^i_k ( t )  dt. \label{Eqn_bar_a}
\end{eqnarray}

The  paper mainly  considers  two agent problem. The $N$   agent problem is discussed  in section \ref{sec_N_player} and the results are conjectured using the two agent results.   {\it Thus for   simpler notations, we represent the tagged customer by $i$ while the other customer is   indexed by $j$.}

 The expected  utility for stage $k$  equals:
\begin{eqnarray}
 \label{Eqn_run_cost}
r_k^i(\z_k, \a_k; \pi^j) \hspace{-10mm} &  \\ &= \left \{ \begin{array}{lll}
P_M^i  1_{k=M} - \nu  E\left [ \ba_k^i (d_k^i )  \right ] & \mbox{ if } \tau_{k-1}^i  < T   \\
0    & \mbox{ else,} 
\end{array}   \right .  
\nonumber
\end{eqnarray}
where $P_M^i $   represents 
the probability of eventual success  (when all the $M$ locks are contacted before $T$ and when the first contact is successful) and $\nu > 0$ is the trade-off factor between the reward and the cost.  Note here that the reward (probability of success)  is added only to the last stage, i.e., only when $k = M$.

For one lock problem, i.e., when $M=1$, the probability of success equals
\begin{eqnarray}
P_1^i  = 
 \int_{0}^{T }  P_f^{j} (s \big |   \pi^{j})  f_{\tau, 1}^i  (s) ds, \label{Eqn_Pki_One_lock}
\end{eqnarray}where $ P_f^{j} (s \big | \pi^{j}) $ is the probability that the other agent has not contacted  the lock before  the agent $i$, i.e., before time $\tau_{1}^i \approx s$ and 
(see equation (\ref{Eqn_bar_a}) for definition of $\ba^i_k (\cdot)$)
\begin{eqnarray}
\label{Eqn_Density}
f_{\tau, 1}^i (s) &:=&   \exp\left (  - \ba^i_1  (s) \right ) a^i_1 ( s)    
\end{eqnarray}
is the density\footnote{It is clear that the complementary CDF of $\tau^i_1$ (time till  agent $i$ contacts the destination, immaterial of whether it is the first one or not and whether it is before $T$ or not), under deterministic policy $\pi^i$, is not influenced by the strategies of others and is given by:
\beq
P(\tau^i_1 > t) &=&  \exp\left (  - \int_0^t a^i_1  (s)  ds \right ) \mbox{ and thus its PDF is given by } \\
f_{\tau, 1}^i (t)  &=& \exp\left (  - \ba^i_1  (t)  \right )a^i_1  (t) \mbox{ with } \ba^i_1  (s)  \ := \int_0^t  a^i_1  (s)  ds.
\eeq} of the  associated contact process. Recall   the contact process for $k$-th contact\footnote{We would like to emphasise here that $k$ represents the number of the contacts.
}, is a Poisson process with time varying rate given by $\a^i_k$.  Note simply that  the probability of   agent $j$ not contacting the first lock before agent $i$, for the given   strategy pairs,  equals (more details in the proof of Theorem \ref{Thm_silent})
$$ P_f^{j} (s \big |  \pi^{j}) = P \left  (  \tau_1^{j} > s | \tau_1^i \approx s \right ) =  \exp\left (  - \ba^{j}_1  (s) \right ).  $$

In a similar way for the two lock problem, 
\begin{eqnarray}
P_2^i  
= \int_{0}^{T}  P_f^{j} (s \big |   \pi^j) \left (   \int_s^T    f_{\tau, 2}^i  (u) du \right )  \ \   f_{\tau, 1}^i  (s) ds, \label{Eqn_Pki}
\end{eqnarray}where $ P_f^{j} (s \big |   \pi^{j}) $ is the probability that the other agent (agent $j$) has not contacted the first lock before agent $i$ (which is the same as the one discussed in the one lock problem) and   
\begin{eqnarray}
\label{Eqn_Density2}
f_{\tau, 2}^i (\tau^i_1 + s) &:=&   \exp\left (  - \ba^i_2  (s) \right ) a^i_2 ( s)    
\end{eqnarray}
is the density of the  second-lock contact process of agent $i$. 

It is easy to observe that for any given $\a \in L^\infty$, the expected cost equals (see (\ref{Eqn_bar_a}) and with $\tau_0^i := 0$):

\vspace{-4mm}
{\small \begin{eqnarray}
\label{Eqn_Cost}
E[\ba_k^i (d_k^i )] &=& 
 E\left [ \int_{0}^{d_{k}^i  }  a_{k}^i (s) ds  \right ] \nonumber \\
 & = & \ba_k^i (T-\tau_{k-1}^i) \exp ( - \ba_k^i (T-\tau_{k-1}^i) )   \nonumber \\
 && \hspace{-3mm}+ \int_0^{\ T-\tau_{k-1}^i }   \ba_k^i (s)      \exp\left (  - \ba^i_k  (s) \right ) a^i_k ( s)  ds.   \hspace{5mm}
\end{eqnarray}}
If the contact occurs  after the deadline $T$, one has to pay for the entire duration $T-\tau_{k-1}^i$  (with zero reward) and hence the first term in the above equation.

{\bf Game Formulation:}  The overall utility of agent $i$, when player $j$ chooses the strategy $\pi_j$   is given by
\begin{eqnarray}
J^i (\pi_i, \pi_j ) =  \sum_{k=1}^M  E[ r_k^i(\z_k, \a_k;  \  \pi_j ) ].
\end{eqnarray}
Thus we have a strategic/normal form non-cooperative game problem and our aim is to find a pair of policies (that depend only upon the available (partial) information)  that form the Nash equilibrium (NE). 

We begin with a one lock problem in the next section. 

\section{One lock problem}
We specialize to one lock problem in this section, while the two lock problem is considered in the next section. Here the agent that first gets the lock, gets the reward. At any time, the agents are aware of their own state, but they would know the information of the others only when it contacts the lock.  At that time, the contact can be successful or unsuccessful, with the former event  implying the other agent has not yet contacted the lock.   In this case the agents have to choose one contact rate function $a^i (\cdot) $/$a^j (\cdot)$  at the start, and would stop either after a contact or after the deadline  $T$ expires.  There is no major information update at any  time point before the (only one) contact and hence this control process is sufficient. 

 We prove that a pair of (time) threshold policies form an NE for this game. We prove this by showing that the best response against a threshold policy is again threshold. 
 Towards this,   we first discuss the   best response against any given strategy of the opponent. 
Given a policy $\pi^j =  \a^j  $  with $\a^j   \in L^\infty [0, T]$ of agent $j$, it is clear that  the failure probability  of the other  agent $j $ in equation (\ref{Eqn_Pki_One_lock})  equals:
$$
P_f^{j} (s \big | \pi^{j})  = \exp(-\ba^j (s )  ),     
$$where $\ba^j (\cdot)$ is as defined in (\ref{Eqn_bar_a}).   The best response for this case is obtained by (see equations (\ref{Eqn_bar_a})-(\ref{Eqn_Cost})):

\vspace{-4mm}
{\small\begin{eqnarray}
v^i_1 (\z_1; \pi_{j})  \hspace{-1mm} &\hspace{-1mm}=\hspace{-1mm}& \hspace{-1mm}  \sup_{\a^i \in L^\infty }  J(\a^i; \a^j) \mbox{ with }  \nonumber \\ 
J(\a^i; \a^j)\hspace{-1mm} &\hspace{-1mm}=\hspace{-1mm}& \hspace{-1mm}  \int_0^T  \left ( \exp(-\ba_1^j (s )  )   - \nu  \ba^i (s)  \right ) \exp (-\ba^i(s) ) a^i ( s) ds    \nonumber \\
&& \label{Eqn_BR_one_lock}
 - \nu \ba^i(T) \exp (-\ba^i(T) ) . 
\end{eqnarray}}For any given policy $\pi^j = \a^j (\cdot)$ of the other agent,   the above  best response is clearly a finite horizon ($T$) optimal control problem as $J$ can be rewritten as
\begin{eqnarray}
\label{Eqn_opt_cntrl_one_lock}
J(\a^i) = \int_0^T \left ( h^j (s)    - \nu   x (s)  \right ) \exp (-x(s) ) a^i ( s) ds   \\  +g(x(T)),  \nonumber
\end{eqnarray}
with state process
$$
\stackrel{\bm\cdot}{x}  (t) =    a^i ( t)  \mbox{ and thus }  x(t) = \int_0^t a^i (t) dt = \ba^i (s),
$$a given function 
$$
h^j(s) := \exp(-\ba^j (s )  ),
$$and
 with terminal cost 
\begin{eqnarray}
\label{Eqn_terminal_cost}
g(x) = - \nu x\exp (-x ) .
\end{eqnarray}
{\it  Here we need to note that $x(t)$ represents the state process for the optimal control problem that is used as a tool  to solve the original best response problem and is not the state process of the actual/original problem.}    Further in two lock problem (considered in the next section), we will have one such optimal control problem for each stage and for each state and each of those optimal control problems will have their own state processes.  

{\bf Conjecture: } We  aim to prove using Hamilton Jacobi  (HJB) equations (which provide solution to the above mentioned optimal control problems of each stage), that the best response against any policy of agent $j$ would be a time-threshold type policy as discussed below.  We are yet to prove this.  However from the nature of the HJB equations one can conclude that the best response policies are of bang-bang type. 
Currently we continue with deriving the best response against  time-threshold policies.

\subsection{Best response against  silent   opponent }
\label{Sec_other_silent}
We  begin with best response of an agent, when the other agent is silent, i.e., when $a^j (t) = 0$ for all $t$. This particular result will be used repeatedly (also for the case with two locks) and hence is stated first. 
Let ${\cal C} : =  \{ a \in  L^\infty  :  ||a|| \le \bnd^i \}$.

\begin{thm}{\bf [Best response, against silent opponent]}
\label{Thm_silent}
When an agent attempts to acquire a lock  for any given time $U$,  and when there is no competition and if it receives a reward  $c$ upon success:   i)
  the  best response  of  (\ref{Eqn_BR_one_lock}) is derived by  solving the following  optimal control problem:
\begin{eqnarray*}
v(x) := \sup_{ a (\cdot) \in  {\cal C} }   \bigg \{  \int_{0}^U   \big ( c- \nu x(s) \big ) \exp ( -x (s)  )  a (s) ds    & \\
& \hspace{-43mm}  - \nu x(U)  \exp (- x(U))  \bigg \} \\
& \hspace{-100mm} \mbox{ with }     \stackrel{\cdot}{ x} (t)  =  a(t) \mbox{ and }   x(0) = x .
\end{eqnarray*}
 ii)  The solution of the above  problem  is the following:
 \begin{eqnarray}
 a^* (t) =  \hspace{-2mm}& \hspace{-2mm} \left \{
 \begin{array}{llll}
 \bnd^i   & \mbox { for all }   t    & \mbox{ if }  \nu \le  c \nonumber \\
 0  & \mbox { for all }   t    & \mbox{ if }  \nu  > c
 \end{array} 
  \right .  \mbox{ and } \hspace{20mm}
   \nonumber  \\
v(x)  
= \hspace{-2mm}& \hspace{-2mm} \left \{\hspace{-2mm}
 \begin{array}{llll}
  \mbox{\small$\left [ (c-\nu)  \left (1- \exp(- \bnd^i U) \right )  - \nu \x    \right ] \exp(-\x) $}     \hspace{-30mm} & \mbox{if }  \nu \le  c \\
  - \nu \x     \exp(-\x)      & \mbox{if }  \nu  > c.
 \end{array}
 \right .   \nonumber  \\ \label{Eqn_Silent_vx}
 \end{eqnarray}
 
\end{thm}
{\bf Proof:} We drop the superscript $i$ in this proof, for simpler notations. Using density (\ref{Eqn_Density}),   the expected reward  (against silent opponent) equals
\begin{eqnarray}
\label{Eqn_Rewar_silent}
  E[R] =  cE[\tau_1  < U ]    
& = & c \int_0^U  \exp ( - \ba  (t) ) \a(t) dt  \nonumber \\
&=& c \int_0^U \exp ( - \x(t) ) \a(t) dt  .   
\end{eqnarray} 
The cost does not depend upon the existence of other players, hence remains the same as  in  (\ref{Eqn_Cost}), reproducing here for clarity:
\beq
E[\ba (d_1) ]
   = 
 \exp (-  x (U) )   \x (U)   +  \int_0^U      \x (t)    \exp\left (  - \x (t) \right ) \a(t)   dt  
 \eeq
Then the overall problem is to maximize $E[R] - \nu E[\ba (d_1) ]$ and hence we consider
\beq
 \sup_{ \a (.) }   \left \{  \int_0^U  L(t, \x (t) ,  \a(t) )  dt   + g(x (U)  ) \right \}  \mbox{ with }  \\
L(t, \x, \a) = 
  \left (    c - \nu x \right ) \exp{(-x)}a  \mbox{ and }  g(\x) = - \nu \x \exp (-\x).
\eeq
Thus we need the solution of the following (Hamiltonian Jacobi) HJB PDE, with $v(t,x)$ representing the value function and with $v_t$, $v_x$ its partial derivatives
\beq
v_t  (t, \x) + \max_{a \in [0, \bnd]}  \left \{ L(t, \x, \a)  + \a v_x (t, \x) \right \} & =& 0\eeq  or in other words  
\beq
v_t  (t, \x) + \max_{a \in [0, \bnd]}  \left \{  \left (    c - \nu \x \right ) \exp{(-\x)} \a    + \a v_x (t, \x) \right \} & =& 0\eeq
with boundary condition 
$
v(U, x) =   g(\x) =  -\nu \x \exp (-\x).
$\\
{\bf Claim:} We claim the following is a solution\footnote{We derived the above solution by solving the HJB PDE after replacing the maximizer with  $\bnd$.} satisfying the above boundary valued problem (when $\nu < c $): 
\beq
W(t, \x) &=&   \left ( - \nu \x - \nu +c  \right )\exp(-\x) + \kappa \exp (-\x + \bnd t )   \mbox{ with} \\
\kappa &=&  \exp (-\bnd U) (\nu - c).
\eeq
Note that its partial derivatives are:
\beq
W_x(t, \x)  &=&  \left ( \nu \x - c  \right )\exp(-\x) - \kappa \exp (-\x + \bnd t ), \\
W_t(t, \x)  &=&   \bnd \kappa \exp (-\x + \bnd t ), \mbox{ and clearly for any } a, \\
 L(t, \x, \a)  + \a W_x (t, \x)  \  = \   - \kappa \exp (-\x + \bnd t )  a. \hspace{-50mm}
\eeq
Thus if $\kappa \le  0$, the maximizer in HJB PDE is $\bnd^i$ and then $W(., .)$ satisfies the HJB PDE, 
\beq
W_t (t, \x)   +  \kappa \exp (-\x + \bnd t )  \bnd  &=& 0, \eeq
 and also satisfies the boundary condition 
 
 \vspace{-6mm}
 {\small \beq
W (T, \x)& =&  \left ( - \nu \x - \nu +c  \right )\exp(-\x)  \\&&+ \exp (-\bnd U) (\nu - c)  \exp (-\x + \bnd U )   \\
&= & - \nu \x \exp(-\x)   \mbox{ for any } x.
 \eeq}
 It is further easy to verify that $\a^* (t) = \bnd^i$ for all $t$ and $\x^* (t) = \bnd^i t$  satisfy equation (5.7) of \cite[Theorem 5.1]{Feller}.
 Thus when $\kappa \le  0$ or equivalently when  $\nu \le  c$ then the optimal policy is to attempt with highest possible rate  all the times. 
 
 When $\nu > c$, using similar logic one can show that  $W(t,x) = -\nu x \exp(-x)$  (for all $x, t$) is the solution of the HJB PDE and  $\a^*(t)  = 0$ for all $t$.
\eop

Using similar techniques one can find  best response against any given   time-threshold policy, which is next considered.

\subsection{ Best response against a time Threshold policy}
Assume now player $j $ uses the following time-threshold policy, represented by:
$$
\Gamma(\psi^j): a^j (t) = \bnd^j  {\cal X}_{[0, \psi^j] } (t) , \mbox{ with }  \psi^j \le  T.
$$Basically agent $j$ attempts with maximum acceleration $\bnd^j$ till time $\psi^j$ and stops completely after that. 
In this case the failure probability  of  agent $j $ in equation (\ref{Eqn_Pki_One_lock})  simplifies to:
$$
P_f^{j} (s \big |  \pi^{j})  = \exp(-\bnd^j (s\wedge \psi^j)  ),     \mbox{ when }  \pi^j = \Gamma(\psi^j),  
$$and so

\vspace{-10mm}
{\small\begin{eqnarray}
\label{Eqn_hj_threshold}\hspace{5mm}
h^j(s) = \left \{  \begin{array}{llll}
\exp (-\bnd^j s )   &\mbox{ if }  s \le \psi^j  \\
\exp (-\bnd^j  \psi^j )   &\mbox{ if }  s >  \psi^j.
\end{array}
\right .
\end{eqnarray}}
The best response against such a Threshold policy of agent $j$ is obtained in the following.
From Theorem \ref{Thm_silent}, it is clear that  the best responses against any strategy would be to remain silent when $\nu \ge 1$ and when the reward equals one. 
Thus the  Nash equilibrium  strategies  would   be to remain silent by both the agents for $\nu \ge 1$. From now on,  we consider    $\nu < 1.$
\begin{thm}
\label{Thm_BR} {\bf [Best response]} Assume $\nu < 1.$
The best response of agent $i$  against  $\Gamma$ ($\psi$) policy of  agent $j$ is given by:
{\small\[
  BR_i (\Gamma(\psi)) = 
  \begin{cases}
    \Gamma(T) & \text{if } \nu < \exp{(-\bnd^j\psi)}  \\
    \Gamma(\theta^i_{\nu}   ) & \text{else.} 
 \end{cases}
\]}
where,   \vspace{-6mm}
$$
\theta^i_{\nu }  =  \min \left \{-\frac{  \ln ( \nu  )  }{\beta^j} , T\right \} .
$$
\end{thm}
{\bf Proof:}  The details of this proof are in Appendix A. \eop
Thus when agent $j$ uses threshold strategy with small $\psi$,  best response of agent $i$ is to attempt  till the end and 
if the threshold of agent $j$ is larger,  $\psi \ge  \theta_\nu^i$ then  the best response of agent $i$ is to try till $\theta_\nu^i$  (irrespective of the actual value of $\psi$).

\vspace{-5mm}

\subsection{Nash Equilibrium}
We observe from the above result that the best response against a threshold policy is again a threshold policy. Thus one can get the Nash equilibrium if one can find  two thresholds one for each agent, such that 
$$
\Gamma (\psi^i )  \in    BR_i (\Gamma(\psi^j))  \mbox{ and }  \Gamma (\psi^j )  \in    BR_j (\Gamma(\psi^i)).
$$From Theorem \ref{Thm_BR}, it is   easy to find such a pair of thresholds  and is also easy to verify  that this  pair of thresholds  is  unique. We have the following:
\begin{thm}\textbf{[Nash Equilibrium]}  Assume $\nu < 1$, and without loss of generality $\beta^j \ge \beta^i$.  For a two agent partial information game, we have a  Nash equilibrium among (time) threshold policies,  as defined below:
\[
\begin{array}{llll}
    \left (\Gamma(\theta^i_{\nu}  ), \   \ 
    \Gamma(\theta^j_{\nu}  ) \right )    &\mbox{ if }  \beta^j = \beta^i \\
     \left (\Gamma(\theta^i_{\nu}  ), \   \ 
    \Gamma( T  )  \right )  &\mbox{ if }  \beta^j >  \beta^i  
\end{array}  
\]where threshold  $\theta^i_{\nu}  $  is as in Theorem \ref{Thm_BR}, while  $\theta^j_{\nu}  $ is given by:
 $$\hspace{15mm}
\theta^j_{\nu} =     \min \left \{-\frac{  \ln   (\nu)  }{\bnd^i } ,   T \right \} . \hspace{15mm} \mbox{ \eop }
$$
\end{thm}
{\bf Proof:} The first line  is  easily evident from Theorem \ref{Thm_BR}. For the second one,  observe the following: when $\psi^i = \theta_\nu^i$:
$$
\exp ( - \beta^i  \theta_\nu^i ) =  \exp (  \beta^i  \ln (\nu)  / \beta^j )  \ge   \nu,     
$$because  $\beta^i  \ln (\nu)/  \beta^j  \ge  \ln (\nu) $  (note $\ln(\nu) < 0$) and thus  
$$
BR_j (\theta_\nu^i) = \Gamma (T) . 
$$ Now if $\exp (-\beta^j T)  \le  \nu$,    then clearly 
$
BR_i ( T) = \Gamma (\theta_\nu^i) . 
$ On the other hand if $\exp (-\beta^j T)  >   \nu$,      then 
$\theta_\nu^i = T$. \eop

It is further clear that  (simple calculations using Theorem \ref{Thm_BR}) we have unique Nash Equilibrium among time threshold policies. It would be more interesting if we can show this is the unique NE, but that would be a part of future work.

Thus when one has no access to the information of the other agent till their own contact with the lock, the NE are given by open loop policies. But this is true only for one lock ($M=1$) problem. With large $M$, we will have closed loop policies but the policies change only at major information change epochs. In all, we will see that the NE will again have a group of open loop policies, each of which is used till a major  change in the information.

\section{Two Lock Problem}

Before we proceed with the analysis we would summarize the protocol again.
{ Any agent succeeds only if it  contacts lock one, followed by lock two and 
only the agent that gets both the locks receives   reward one.  
If a particular agent contacts the lock one,  we say it had an unsuccessful contact if it is not the first one to contact the lock. If an agent's contact is unsuccessful,   there is no incentive for the agent  to try any further.  On the other hand when an agent  is successful, it knows it would be the only one to   chase the second lock. We can use the previous ($M=1$ case) analysis, Theorem \ref{Thm_silent}, to compute  the best response against silence  opponent (for second lock).}

This is a two stage problem, as the utility of  agent $i$ is given by:
\begin{eqnarray*}
J^i (\pi_i, \pi_j ) =  \sum_{k=1}^2  E[ r_k^i(\z_k, \a_k;  \  \pi_ j ) ].
\end{eqnarray*}
For this two lock case,  the best response  of agent $i$ against any given strategy of  player $j$ can be solved using (two stage) dynamic programming (DP) equations as below

\vspace{-4mm}
{\small \begin{eqnarray}
v^i_{k} (\z_k; \pi_{j} )\hspace{-1mm}  &\hspace{-1mm}  = \hspace{-1mm}  & \hspace{-1mm} 0 \mbox{ if }  k = 3 \mbox{ or if } \tau_{k-1}^i >  T \mbox{, and else,}  \label{Eqn_DPs}   \\
v^i_k (\z_k; \pi_{j}) \hspace{-1mm}  &\hspace{-1mm}  = \hspace{-1mm}  & \hspace{-1mm} 
 \sup_{\a_k}  \left \{ r_k^i (\z_k, \a_k; \pi_{j}) + E[  v^i_{k+1} (\z_{k+1}; \pi_{j}) ) | \z_k, \a_k] \right \}  , \nonumber 
\end{eqnarray}}with stage wise costs as defined in equations (\ref{Eqn_run_cost})-(\ref{Eqn_Cost}).
Note these DPs hold even when the action spaces are Banach spaces, as in our case.  

Like in one-lock case, we obtain a NE, by finding best response against appropriate threshold strategies. 

{\bf  Threshold  strategy for two-lock problem:} 
Our conjecture is that the strategy constructed using state dependent time-threshold policies will form a part of the NE.   At  contact instance of the first lock, the contact could be successful or unsuccessful. Thus we have two types of states immediately after the first contact, i.e., the state after the first contact is either given by  $\z_2 = (s, \tau)$ or by  $\z_2 = (u, \tau)$.   We compactly represent Threshold policy by $\Gamma_2 (\psi)$ which means the following:
\begin{eqnarray*}
\mbox{at start, use }   &\Gamma (\psi) \mbox{ policy, }  \\
\mbox{if } \z_2 = (s, \tau)  \mbox{, i.e., when successful use } &\Gamma (T-\tau) \mbox{   and  }  \\
\mbox{if } \z_2 = (u,  \tau)  \mbox{, i.e., when unsuccessful use } &\Gamma (0) \mbox{ policy. }  
\end{eqnarray*}
Theorem \ref{Thm_silent}, inspires us to conjecture that   this kind of a threshold strategy becomes a part of the NE and the same is proved in Theorem \ref{Thm_NE_two_lock}.
We begin with the best response. 
 
\subsection{Best response against a  Threshold strategy}  Say agent $j$ uses threshold strategy $\Gamma_2 (\psi^j)$.  We obtain the best response by solving the DP equations (\ref{Eqn_DPs}) using backward induction.  When $k = 2$ in (\ref{Eqn_DPs})  and if $z_2 = (u, \tau)$  it is immediately clear that  (see (\ref{Eqn_run_cost}))
$$
v_2 (\z_2;  \Gamma_2) = 0 \mbox{ for any } \tau,
$$as failure with first lock implies zero reward. 
If $\z_2 = (s, \tau)$, i.e., if the player $i$ is successful with first lock and the contact was at $\tau$, the  agent $j$ will either have  unsuccessful  contact or may not even contact the first lock before the deadline $T$.   Further because agent $j$ uses $\Gamma_2 (\psi)$ policy   it would not try for the second lock. Thus agent $i$ will attempt for second lock, while the other agent is silent  with respect to second lock.   Thus the optimization problem corresponding to this stage from equations (\ref{Eqn_run_cost})-(\ref{Eqn_Cost})   is given by: 
\begin{eqnarray*}
  \sup_{ a (\cdot) \in  {\cal C} }   \bigg \{  \int_{0}^U   \big ( 1- \nu x(s) \big ) \exp ( -x (s)  )  a (s) ds    & \\
& \hspace{-53mm}  - \nu x(U)  \exp (- x(U))  \bigg \} \\
& \hspace{-60mm} \mbox{ with }     \stackrel{\cdot}{ x} (t)  =  a(t) \mbox{ and }   x(0) = x , \mbox{ with }  U = T- \tau.
\end{eqnarray*}
This is exactly the optimization problem considered in Theorem \ref{Thm_silent}  with $U = T-\tau$ and hence the best response (with $\nu < 1$) is given by:  $\Gamma (T-\tau)$  (attempt with maximum for the rest of the  period). 
Thus from Theorem \ref{Thm_silent}  with  $U = T- \tau$ and $x  = 0$ we  have:
\beq
v_2(s, \tau; \Gamma_2)  &= &    v(0  )  
=  (1-\nu)  \left (1- \exp(- \bnd^i  (T-\tau)  ) \right )     \mbox{ and}  \\
v_2 (u, \tau; \Gamma_2)& =& 0.
\eeq
 Now solving the DP equations for $k = 1$, it is easy to verify that the corresponding optimization problem is (with $x(\cdot)$ as before and see (\ref{Eqn_run_cost}), (\ref{Eqn_DPs})):

\vspace{-4mm}
{\small \beq
\sup_{a (\cdot) } \bigg \{  - \nu    \int_0^T    x  (\tau)    \exp(- x (\tau) ) a (\tau) d\tau  - \nu x  (T)  \exp (-x(T) )   \hspace{-80mm} &
\\ 
& \hspace{13mm}
+    \int_0^T   \exp(-\ba^j (\tau) )  v_2(s, \tau; \Gamma_2)    \exp(-x (\tau) ) a (\tau) d\tau \bigg \}.
\eeq}
This optimization problem is once again solved using optimal control theory based tools and we directly obtain the following. When $\nu \ge  1/2$ it is easy to verify that, both agents  being silent  is the Nash equilibrium. This result can easily be derived (by finding the best responses as in Theorem \ref{Thm_Silence_twolock} of Appendix B, which provides the best response against the silent opponent).

\ignore{
\begin{thm}{\bf[Best Response against threshold]}
\label{Thm_BR_two_lock}
When
$\nu < \  1/2, $ the best response against the thresholds policies  $\Gamma(\psi^j, T, 0)$ is given by:
\begin{eqnarray*} 
 BR^i ( \Gamma_2(\psi^j) )  & =& \Gamma_2 (T )    
\end{eqnarray*}
\end{thm}
{\bf Proof:}  Assume agent $j$ plays the threshold policy $\Gamma_2 (\psi^j)$.  We find the best response of agent $i$, as before, by first finding the best response for the last $T-\psi^j$  amount of time,  and then using the value function at $\psi^j$ as the terminal one for the first part which spans over $[0, \psi^j)$. During the last leg, the reward which equals the  probability  that agent $i$ wins is given by  $c = \exp (-\bnd^j \psi^j)$. Then the best response (for time period between $[\psi^j, T]$ is given by Silence theorem \ref{Thm_Silence_twolock} (given in Appendix B)  with $U - T -\psi^j$.

Now the proof is complete if we solve the following optimal control problem
\beq
\sup_{a} \bigg \{
\int_0^{\psi^j} \left(1-\nu \right)\exp (-\bnd^jt)\exp(-x) \\  - \left(1-\nu \right)\exp (-\bnd T)\exp (-\bnd^jt)\exp (\bnd^it)\exp(-x)-\nu x\exp(-x) \bigg \}
\eeq with terminal as given by Theorem \ref{Thm_Silence_twolock}
The rest of the proof is in Appendix B.   \eop

Thus we are now in a position to derive the NE and that is achieve in the following.  }

\begin{thm}{\bf[Nash Equilibrium]}
\label{Thm_NE_two_lock}Let $\nu < 1/2$ and assume $\beta^j \ge \beta^i$.  The NE is given by the following, under the conditions:

\vspace{-5mm}
{\small\begin{eqnarray*}
\begin{array}{lll}
  \bigg ( \Gamma_2 (0), \Gamma_2( 0 ) \bigg ),  
  & & \hspace{-45mm}\mbox{if }   \ \exp (-\beta^j T)  >  \frac{1- 2\nu}{1- \nu  } ,    \\ 
\bigg ( \Gamma_2 (0), \Gamma_2(\psi_0^{j*}) \bigg ),  & \mbox{with }   \psi_0^{j*} = T + \frac{1}{\beta^j}  \ln \left  (\frac{1- 2\nu}{1- \nu} \right )  \\
  & & \hspace{-45mm}\mbox{if }   \exp (-\beta^i T) >  \frac{1- 2\nu}{1- \nu  } > \exp (-\beta^j T) .  
\end{array}
\end{eqnarray*}}
The above $\psi_0^{j*} > 0$. 
Now consider that $\exp (-\beta^i T) \le  \frac{1- 2\nu}{1- \nu  }.$
Let $\psi^{i*}$ satisfy  

\vspace{-4mm}
{\small
 \begin{eqnarray*}
\exp (-\bnd^j \psi^{i*} )  = \min   \left \{1,   \exp (-\bnd^i (T-\psi^{i*}) -\bnd^j  \psi^{i*} )   +  \frac{ \nu}{ \left(1-\nu \right)}  \right \}.
 \end{eqnarray*}}
and  $\psi^{j*}$ satisfy the following equation 

\vspace{-4mm}
{\small \beq
  \exp (-\bnd^i \psi^{j*} )  =    \min   \left \{1,  \exp (-\bnd^j (T-\psi^{j*}) -\bnd^i \psi^{j*})   + \frac{ \nu }{1-\nu} \right \}.
\eeq}
 If the following two conditions are satisfied

\vspace{-4mm}
{\small \begin{eqnarray}
\label{Eqn_conditions_for_NE}
 \exp(- \bnd^i (T-  \psi^{j*}) )  &>&   \frac{\exp(-\bnd^j \psi^{j*}) - 2\nu }{\exp(-\bnd^j \psi^{j*}) - \nu}    \mbox{ and }  \nonumber \\
 \exp(- \bnd^j (T-  \psi^{i*}) )  &> &  \frac{\exp(-\bnd^i \psi^{i*}) - 2\nu }{\exp(-\bnd^i \psi^{i*}) - \nu}     ,
\end{eqnarray}}
 then  the pair   $\bigg ( \Gamma_2 (\psi^{i*}), \Gamma_2(\psi^{j*}) \bigg )$ forms a Nash equilibrium. 
The above $\psi^{i*}, \psi^{j*} < T$.
   The   conditions (\ref{Eqn_conditions_for_NE}) are immediately satisfied with $\bnd^j = \bnd^i = \bnd$, in which case the common
 $$ 
 \exp (-\bnd \psi^{*})  =   \min   \left \{1,  \   \exp (-\bnd T)   + \frac{ \nu }{1-\nu} \right \}.    \mbox{ \eop }
 $$ 
\end{thm}
 {\bf Remarks:} Few interesting observations for the cases that we derived the result:  a) in two lock problem none of the agents at an NE would try till $T$ (in contrast to one lock problem); b) the agents either remain silent or attempt for a time period that is strictly less than $T$; and c) we obtained NE for all the values of the parameters for the case when $\beta^j = \beta^i.$

\section{ Extensions and Future work}
\label{sec_N_player}
One can easily extend the results to $N$-player game with symmetric parameters, i.e., to the case when $\beta^i = \beta$ for all $i$. For one lock problem it is not difficult to conjecture that the Nash equilibrium among time-Threshold policies is given by, 
\beq
\bigg ( \Gamma (\theta^*), \Gamma (\theta^*), \cdots, \Gamma (\theta^*) \bigg ) \mbox{ with }  \\
\theta^* := \left \{ \begin{array}{lllll}
   -\frac{  \ln ( \nu ) } { (N-1) \beta }   & \mbox{ if }   \exp(-\beta (N-1) T)  \le  \nu\\
   T   &\mbox{ else.}
\end{array}      \right .
\eeq
In a similar way the two-lock Nash equilibrium for symmetric agents, could probably be obtained  using Theorem \ref{Thm_NE_two_lock}; with the parameter of the opponent as $\beta^j = (N-1)\beta$  and with $\beta^i = \beta$. We conjecture the NE for this case to be,  $\bigg ( \Gamma_2(\psi^{i*}), \Gamma_2(\psi^{i*}),  \cdots, \Gamma_2(\psi^{i*}) \bigg)$. 
These are only conjectures and we need to verify and prove the same. Further we would like to work with asymmetric agents. 

It would be equally interesting to work with $M$-lock problem with $M >2$. We anticipate that the silence theorem (like Theorems \ref{Thm_silent} and \ref{Thm_Silence_twolock}) should be extended and then the analysis would follow easily. It would be more interesting to work with the problem in which each lock fetches a reward.  For all these and more general problems,  the methodology would be the same;  One needs to consider open loop control till a new information update.  Thus these partial information problems would span from completely open loop policies  (no information) to  completely closed loop policies (or full information). 
\vspace{-7mm}
\section*{Conclusions}
We considered lock acquisition games with partial, asymmetric information. Agents attempt to control the rate of their Poisson clocks to acquire two locks, the first one to get both would get the reward.  There is a deadline before which the locks are to be acquired,  only  the first agent to contact the lock can acquire it and the agents are not aware of the acquisition status of others. It is possible that an agent continues its acquisition attempts,  while  the lock is already acquired by another agent.  The agents pay a cost proportional to their rates of acquisition. 
We proposed a new approach to solve these asymmetric and non-classical information games, "open loop control till the information update".  
With this approach we have dynamic programming equations applicable at state change update instances and then each stage of the dynamic programming equations is to be solved by optimal control theory based tools (HJB equations). 
We showed that a pair of (available) state dependent time threshold policies form Nash equilibrium. We also conjectured the results for the games with $N$-agents.


\section*{Appendix A: Proofs related to One Lock Problem}

{\bf Proof of Theorem \ref{Thm_BR}:}  
The best response against a threshold policy  can be obtained by solving  the optimal control problem  (see equation (\ref{Eqn_opt_cntrl_one_lock}) with $h^j$ as in (\ref{Eqn_hj_threshold}))
\begin{eqnarray*}
v( x) := \sup_{ a (\cdot) \in  {\cal C} }   \bigg \{  \int_{0}^T  L(t, x(t), a(t) ) ds   - \nu x(T)  \exp (- x(T)) \bigg \}    & \\
& \hspace{-130mm} \mbox{ with state update equation given by}  \\
& \hspace{-110mm}  \stackrel{\cdot}{ x} (t)  =  a(t) ;   x(0) = x  \mbox{ and
with   running cost, } \\
 \mbox{\small $ L(t, x, a) = 
  \begin{cases}
  \bigg(  \exp{(-\bnd^j t   ) - \nu x \bigg )}\exp{(-x)}a & \text{for } t \leq \psi \\
   \bigg  ( \exp{(-\bnd^j \psi) - \nu x \bigg )}\exp{(-x)}a & \text{else.}
    
  \end{cases}$}
\end{eqnarray*}
Further the 
terminal cost is  $g(x)= -\nu x\exp{(-x)}$.
Thus the 
 HJB (PDE) equation  that needs to be solved as in the proof of Theorem \ref{Thm_silent} is given by the following:
\begin{equation}
\label{Eqn_PDE_2}
    \frac{\partial}{\partial t}v(t,x) + \sup_{a\in [0,\bnd^i]} \left \{L(t,x,a) + a\frac{\partial v}{\partial x}\right \} = 0,
\end{equation}
$$
v(T,x)=-\nu x\exp{(-x)}.
$$
Let $v_t :=  \frac{\partial v}{\partial t}$ and $v_x := \frac{\partial v}{\partial x}$.
 We  conjecture that the optimal control for this problem is a threshold policy $\Gamma(t_1)$ for some appropriate $0\le t_1 \le T$.
 
\textbf{Claim}:We further claim  the following to be the  solution of the above PDE\footnote{We compute the following solutions, replacing the maximizers in HJB PDEs $a^* = \beta^i$. One of them is for the case when $t_1 \le \psi$ and one for the other case.}, we prove this claim alongside computing $t_1$ (we would actually show that $t_1 = \theta_\nu^i$ or $T$):

\vspace{-4mm}
{\small\[
  W(t,x) = \hspace{-1mm}
  \begin{cases}
    -\nu x\exp{(-x)} - \nu\exp{(-x)}  + \frac{\bnd^i }{\bnd^i +\bnd^j }\exp{(-x-\bnd^j t)} \\
    \hspace{1mm}+
     \kappa_1\exp(-x+\bnd^i t) &\hspace{-20mm} \text{if }  t\leq ( t_1 \wedge \psi) \\
       
      - ( \exp (-\bnd^j \psi) - \nu  )   \exp (-\bnd^i t_1)  \exp(-x+\bnd^i t) 
      \\ + ( \exp (-\bnd^j \psi) - \nu  x - \nu)  \exp (-x)  &\hspace{-20mm} \text{if }  \psi \le t\leq t_1\\ 
     
    - \nu x \exp (-x) & \hspace{-20mm}\text{else.} 
    
  \end{cases}
\]}
where,

\vspace{-15mm}
{\small $$
\kappa_1 = \left \{ 
\begin{array}{llll}
 \nu\exp{(- \bnd^i  t_1 )} - \frac{\bnd^i }{\bnd^i +\bnd^j }\exp{(-(\bnd^i +\bnd^j )t_1) }\\
  &\hspace{-10mm}\mbox{ if  }  \nu \ge  \exp (-\psi \beta^j) \\
 \nu\exp{(- \bnd^i  t_1 )} - \frac{\bnd^i }{\bnd^i +\bnd^j }\exp{(-(\bnd^i +\bnd^j ) \psi ) } \\
 + \exp (-\bnd^j \psi)   \bigg(  \exp (-\bnd^i  \psi ) - \exp (-\bnd^i t_1 )    \bigg )  & \mbox{ else. }
\end{array} 
\right . $$}
The partial derivatives of the above are:

\vspace{-4mm}
{\footnotesize{
\[
  W_x(t,x) = 
  \begin{cases}
    \nu x\exp{(-x)}  - \frac{\bnd^i }{\bnd^i +\bnd^j }\exp{(-x-\bnd^j t)}  \\ \hspace{6mm}-\kappa_1\exp(-x+\bnd^i t) & \hspace{-0.20cm}\text{if } t\leq  t_1 \wedge \psi \\ 
    
     + ( \exp (-\bnd^j \psi) - \nu  )   \exp (-\bnd^i t_1)  \exp(-x+\bnd^i t) 
      \\ - ( \exp (-\bnd^j \psi) - \nu  x )  \exp (-x)  & \hspace{-3mm} \text{if }  \psi \le t\leq t_1\\

    (\nu x - \nu) \exp (-x) & \text{else.} 
    
  \end{cases}
\]
\[
  W_t(t,x) = 
  \begin{cases}
     -\frac{\bnd^i \bnd^j }{\bnd^i +\bnd^j }\exp{(-x-\bnd^j t)} + \bnd^i \kappa_1\exp(-x+\bnd^i t) \\  & \hspace{-10mm} \text{if } t\leq t_1 \wedge \psi\\ 

  \\ -  \beta^i ( \exp (-\bnd^j \psi) - \nu  )   \exp (-\bnd^i t_1)  \exp(-x+\bnd^i t)  \\ 
         &\hspace{-10mm} \text{if }  \psi \le t\leq t_1\\

    0 & \text{else.} 
    
  \end{cases}
\]
}}
\TR{
We verify that the above partial derivatives verify  PDE~(\ref{Eqn_PDE_2}), when $t_1$ is set equal to $\theta_\nu^i  $ (defined in the hypothesis of the theorem) and the details are in 
\cite{TR}.}{
Now to check  if the above partial derivatives verify  PDE~(\ref{Eqn_PDE_2}). 
\\{\bf Case 1:  When  $\bm  {\exp(-\beta^j \psi )  \le \nu }$} We will prove for this case that $t_1 = \theta_\nu^i \le \psi$. 
Thus for all  $t \le t_1 \le \psi $ we have: 

\vspace{-4mm}
{\scriptsize
\begin{eqnarray*}
&&\hspace{-10mm}L(t,x,a) + a W_x(t,x)\\ &=& \frac{\bnd^j }{\bnd^i +\bnd^j }\exp{(-x-\bnd^j t)}a - \kappa_1\exp(-x+\bnd^i t )a\\
& =& \bigg(\frac{\bnd^j  }{\bnd^i +\bnd^j }\exp{(-x-\bnd^j (t \wedge \psi)  )}  \\ &&+ 
 \frac{\bnd^i }{\bnd^i +\bnd^j }\exp{(-(\bnd^i +\bnd^j )t_1 ) } \exp{(-x+\bnd^i t)} \\&& -\nu\exp{(- \bnd^i t_1)}\exp(-x+\bnd^i t)\bigg)a\\
& =& \exp{(-x)}\bigg(\frac{\bnd^j }{\bnd^i +\bnd^j }\exp{(-\bnd^j (t \wedge \psi)  )}  \exp(-\bnd^i t) \\&& + \frac{\bnd^i }{\bnd^i +\bnd^j } \exp{(-(\bnd^i +\bnd^j )t_1))} 
\\ && - \nu\exp{(- \bnd^i t_1)}\bigg)a \exp(\bnd^i t).
\end{eqnarray*}
}
For $a^*=\bnd^i$ in this range  we will  require $L(t,x,a) + aW_x(t,x) \geq 0$ and this is true if  for all $t \le t_1 \wedge \psi$

\vspace{-4mm}
{\small\begin{align*}
     \frac{\bnd^j }{\bnd^i +\bnd^j }\exp{(- \bnd^j t   - \bnd^i t ) } &\\  & \hspace{-20mm}
      \geq  \exp{(-\bnd^i t_1)}\bigg(\nu - \frac{\bnd^i }{\bnd^i +\bnd^j }\exp{(-\bnd^j t_1)}\bigg).
\end{align*}}
Thus by monotonicity, the required inequality holds $\forall$ $t\leq t_1$, if it holds at $t=t_1$ and this happens if 
\begin{eqnarray}
\label{Eqn_nu_less}
 \frac{\bnd^j }{\bnd^i +\bnd^j }\exp{(- \bnd^j  t_1 )    } \geq   \bigg(\nu - \frac{\bnd^i }{\bnd^i +\bnd^j }\exp{(-\bnd^j t_1)}\bigg).
\end{eqnarray} 
For $t\ge t_1$ we have
$$
\frac{ L(t, x, a)  + a W_x (t, x) }{a}  = \left ( \exp (- \beta^j t ) - \nu  \right )\exp (-x) 
$$
Thus we will have
  $L(t, x, a)  + a W_x (t, x)\leq 0$ and then $a^* = 0$ for all $t \ge t_1$, if  (by monotonicity)  
  $$\nu \geq \exp(-\bnd^j  t_1 ).$$

  Thus in this case one can set $t_1 = \theta_\nu^i$, i.e., such that  $\exp(-\bnd^j  t_1 ) = \nu $ and this will satisfy all the required conditions, like $W(\cdot, \cdot)$ satisfies the HJB PDE, the boundary condition and also that $a^* = \beta^i$ for all $t\le t_1$ and $a^* = 0$ for all $t \ge t_1$; thus  by  (\cite[Theorem 5.1]{Feller}) 
  $$
  BR_i (\Gamma(\psi) ) = \Gamma (\theta_\nu^i).
  $$


{\bf Case 2:  When  $\bm  {\exp(-\beta^j \psi ) >  \nu }$}  In this case a $t_1 > \psi$ would be the required threshold, in fact  we will see that $t_1 = T$ satisfies all the required conditions.  

We will begin with $t \le \psi$  (with $\psi \le t_1 $) and for this range of $t$ we have:  

\vspace{-4mm}
{\small
\begin{eqnarray*}
&&\hspace{-10mm}L(t,x,a) + a W_x(t,x)\\ &=& \frac{\bnd^j }{\bnd^i +\bnd^j }\exp{(-x-\bnd^j t)}a - \kappa_1\exp(-x+\bnd^i   \psi )a\\
& =& \bigg(\frac{\bnd^j  }{\bnd^i +\bnd^j }\exp{(-x-\bnd^j t  )}  \\ &&+ 
 \frac{\bnd^i }{\bnd^i +\bnd^j }\exp{(-(\bnd^i +\bnd^j )\psi) } \exp{(-x+\bnd^i t)} \\&& -\nu\exp{(- \bnd^i t_1)}\exp(-x+\bnd^i t)
\\ 
 &&  \hspace{-6mm}- \exp (-\bnd^j \psi)   \big(  \exp (-\bnd^i  \psi ) - \exp (-\bnd^i t_1 ) \big )    \exp(-x+\bnd^i t)
 \bigg)a\\
& =& \exp{(-x)}\bigg(\frac{\bnd^j }{\bnd^i +\bnd^j }\exp{(-\bnd^j t   )}  \exp(-\bnd^i t) \\&& + \frac{\bnd^i }{\bnd^i +\bnd^j } \exp{(-(\bnd^i +\bnd^j )\psi))} 
\\ && - \nu\exp{(- \bnd^i t_1)} \\
&&- \exp (-\bnd^j \psi)   \big(  \exp (-\bnd^i  \psi ) - \exp (-\bnd^i t_1 ) \big ) 
\bigg)a \exp(\bnd^i t).
\end{eqnarray*}
}
For obtaining $a^*=\bnd^i$ in this range  we will  require $L(t,x,a) + aW_x(t,x) \geq 0$ and this is true for all $t \le  \psi$ if the following is positive:,

\vspace{-4mm}
{\small\begin{eqnarray*}
\frac{\bnd^j }{\bnd^i +\bnd^j }\exp{(-\bnd^j t   )}  \exp(-\bnd^i t)  \hspace{-2mm} &\ge & \hspace{-2mm}- \frac{\bnd^i }{\bnd^i +\bnd^j } \exp{(-(\bnd^i +\bnd^j )\psi))} 
\\ && \hspace{-30mm} + \nu\exp{(- \bnd^i t_1)} \\
&& \hspace{-30mm} +\exp (-\bnd^j \psi)   \big(  \exp (-\bnd^i  \psi ) - \exp (-\bnd^i t_1 ) \big )   
\end{eqnarray*}}
 
Thus by monotonicity, the required inequality holds $\forall$ $t\leq  \psi$, if it holds at $\psi$ and this happens if 

\vspace{-4mm}
{\small\begin{eqnarray*}
 \exp{(- \bnd^j  \psi - \bnd^i \psi )    }  \\ 
& \hspace{-25mm} \geq   \bigg(\nu \exp{(- \bnd^i t_1   )}  +  \exp (-\bnd^j \psi)      \big(  \exp (-\bnd^i  \psi ) - \exp (-\bnd^i t_1 ) \big )   \bigg).
\end{eqnarray*}}
Or equivalently we require

\vspace{-4mm}
{\small\begin{eqnarray*}
 0 \geq   \bigg(\nu    - \exp (-\bnd^j \psi)        \bigg)   \exp (-\bnd^i t_1 ).
\end{eqnarray*}}
With $   {\exp(-\beta^j \psi ) >  \nu }$, the above is immediate.

Now lets consider 
 \underline{ $\psi \le t \le t_1$}   and as we already mentioned we would like to show that $t_1 = T$ satisfies the required conditions ($L(t,x,a) + aW_x(t,x) \geq 0$ for all $t \ge \psi$ also). In this range of $t$:

\vspace{-4mm}
{\small
\begin{eqnarray*}
&&\hspace{-15mm}L(t,x,a) + a W_x(t,x)\\ &=& 
( \exp (-\bnd^j \psi) - \nu  )   \exp (-\bnd^i t_1)  \exp(-x+\bnd^i t)  a.
\end{eqnarray*}}
Now it is clear that the above is  greater than $0$ for all $t \ge \psi$ as for this case  $   {\exp(-\beta^j \psi ) >  \nu }$ and thus one can set $t_1 = T$.
%
}
\TR{In all,}{

Thus in all,}  one can verify that the following values of $t_1$ satisfy all the required conditions (\cite[Theorem 5.1]{Feller}) and we will have the best response as $\Gamma(t_1)$ with:
\begin{eqnarray*}
\hspace{25mm} t_1 = \left \{ \begin{array}{llll}
\theta_\nu^i   & \mbox{ if }   \exp (- \beta^j \psi  )  \le \nu \\
T  & \mbox{ else. }   \hspace{35mm}  \mbox{\eop}
\end{array} \right . 
\end{eqnarray*}

\ignore{
{\bf Case 1, $\bm {\theta^i_\nu \le  \psi}$:}
Say attempt  with $t_1 \le \psi$.
Clearly for $t\leq t_1$,
{\small
\begin{eqnarray*}
&&\hspace{-10mm}L(t,x,a) + a W_x(t,x)\\ &=& \frac{\bnd^j }{\bnd^i +\bnd^j }\exp{(-x-\bnd^j t)}a - \kappa_1\exp(-x+\bnd^i t)a\\
& =& \bigg(\frac{\bnd^j  }{\bnd^i +\bnd^j }\exp{(-x-\bnd^j t)}  \\ &&+ 
 \frac{\bnd^i }{\bnd^i +\bnd^j }\exp{(-(\bnd^i +\bnd^j )t_1 ) } \exp{(-x+\bnd^i t)} \\&& -\nu\exp{(- \bnd^i t_1)}\exp(-x+\bnd^i t)\bigg)a\\
& =& \exp{(-x)}\bigg(\frac{\bnd^j }{\bnd^i +\bnd^j }\exp{(-\bnd^j t)}  \exp(-\bnd^i t) \\&& + \frac{\bnd^i }{\bnd^i +\bnd^j } \exp{(-(\bnd^i +\bnd^j )t_1))} 
\\ && - \nu\exp{(- \bnd^i t_1)}\bigg)a \exp(\bnd^i t).
\end{eqnarray*}
}
For $a^*=\bnd^i$ in this range  we will  require $L(t,x,a) + aW_x(t,x) \geq 0$ and this is true if 
{\footnotesize\begin{align*}
     \frac{\bnd^j }{\bnd^i +\bnd^j }\exp{(-(\bnd^i +\bnd^j )t))} \geq  \exp{(-\bnd^i t_1)}\bigg(\nu - \frac{\bnd^i }{\bnd^i +\bnd^j }\exp{(-\bnd^j t_1)}\bigg).
\end{align*}}
Thus, the required inequality holds $\forall$ $t\leq t_1$, if it holds at $t=t_1$ and this happens if 
\begin{eqnarray}
\label{Eqn_nu_less}
\nu \leq  \exp(-\bnd^j t_1).
\end{eqnarray}
For $t\ge t_1$ we have
$$
\frac{ L(t, x, a)  + a W_x (t, x) }{a}  = \left ( \exp (- \beta^j (t \wedge \psi) ) - \nu  \right )\exp (-x) 
$$
Thus we will have
  $L(t, x, a)  + a W_x (t, x)\leq 0$ and then $a^* = 0$ for all $t \ge t_1$, if  (by monotonicity)  $\nu \geq \exp(-\bnd^j t_1)$.  
  
  Thus if  $t_1 = \theta_\nu^i :=   \frac{ \ln (\nu) }{-\beta^j}  $, then  it satisfies   $\nu = \exp (-\bnd^j  t_1)$ and $t_1 \le \psi$, and  then the threshold policy $\Gamma(\theta_\nu^i)$   forms the best response against $\pi^j = \Gamma (\psi)$: as  the maximizer in HJB PDE is $\bnd^i$  for all $t \leq t_1$ (because $ L(t, x, a)  + a W_x (t, x)  \ge  0$ for all such $t$)  and 0 for the rest  (because $ L(t, x, a)  + a W_x (t, x)  \leq  0$ for all  $t \geq  t_1$).


When $\frac{ \ln (\nu) }{-\beta^j} > T$, then clearly $\exp (-\beta^j t)  > \nu$ for all $t \le T$ and hence $t_1 = T = \theta_\nu^i$.   

{\bf Case 2,  $\bm {\theta^i_\nu > \psi}$:}  For the case with $\theta_\nu > \psi$, i.e., with $\nu < \exp (-\beta^j \psi)$,  we claim that $t_1 = T$ or in other words $a^* \equiv \bnd^i$, as  in this case for any $t$:

\vspace{-4mm}
{\footnotesize
\begin{eqnarray*}
&& \hspace{-10mm } L(t,x,a)+ a W_x(t,x)\\ &=&\frac{\bnd^j }{\bnd^i +\bnd^j }\exp{(-x-\bnd^j (t\wedge\psi))}a - \kappa_1\exp(-x+\bnd^i t)a\\
&=& \exp{(-x)}\bigg(\frac{\bnd^j }{\bnd^i +\bnd^j }\exp{(-\bnd^j t)} \\ && + \frac{\bnd^i }{\bnd^i +\bnd^j }\exp{(-(\bnd^i +\bnd^j )t_1)}\exp{(\bnd^i t)} \\ && - \nu\exp{(- \bnd^i t_1)}\exp(\bnd^i t)\bigg)a\\&& +  1_{t \ge \psi}\exp{(-x)} \bigg (\exp{(-\bnd^j \psi)} - \exp(-\bnd^j t) \bigg )a.
\end{eqnarray*}}
The last term is always positive, and the rest of the terms  are  similar to the  ones considered above equation  (\ref{Eqn_nu_less}) and hence  $ L(t,x,a)  + a \overline{}W_x(t,x) > 0 $  
for all $t$, as  $\nu < \exp (-\beta^j \psi)$.
For this case,
it is possible that $- \ln (\nu) / \beta^j > T$ and in that case it is clear that one needs to set $\theta_\nu^i = T$.
 \eop
}

%

\section*{Appendix B: Two lock proofs}

\TR{These proofs are provided in \cite{TR}. \end{document} 
}{}

\begin{thm}
\label{Thm_Silence_twolock}{\bf[Best response against Silence with two locks]} 
If the other player is silent and if the player $i$, has to choose an optimal policy that gets reward $c  $, only after contacting both the locks and if this trial is for time $U$, then the corresponding optimal control problem is given by:
{\small\begin{align*}
&\sup_{ a (\cdot) \in  {\cal C} }   \bigg \{\int_0^U   \left(c-\nu \right)\left( 1-\exp(-\bnd^i(U-s))\right)\exp (- x (s) )  a (s) ds  \\& -\nu  \left (  \int_0^{U}   x(s)  \exp(-x(s) ) a (s) ds  + x (U) \exp(-x (U) )   \right )\bigg \}\\
&\mbox{ with }     \stackrel{\bm \cdot}{ x} (t)  =  \a(t) \mbox{ and }   x(0) = x .
\end{align*}}
 ii)  The solution of the above  problem  is the following:
 \begin{eqnarray}
 a^* (t) &=&  \Gamma (\theta^i_{\nu_2} )  \mbox{ with }  \\
 \theta^i_{\nu_2} & :=& \max \left \{ 0,  U +  \frac{1}{\bnd^i} \log  \left ( \frac{c - 2\nu }{c - \nu}  \right )  \right \} 1_{ c > 2 \nu}
   \nonumber 
   \end{eqnarray}
   and the value function is given by:
   
   \vspace{-4mm}
 {\small  \begin{eqnarray}
v(x)  = \left \{\hspace{-1mm}
 \begin{array}{llll}
  (c - 2 \nu) \exp(-x)  \left (1-  \exp (-\bnd^i \theta_{\nu_2} )  \right )    \\  \hspace{-1mm}
   -\left (\nu x   +   ( c-\nu) \bnd^i \theta_{\nu_2} \exp( - \bnd^i U)  \right ) \exp(-x)    \\ 
 &\hspace{-8mm}   \mbox{if }  \theta^i_{\nu_2} > 0   
  \\
  - \nu \x     \exp(-\x)      & \hspace{-8mm}  \mbox{if }  \theta^i_{\nu_2}  = 0.
 \end{array}
 \right . \nonumber \end{eqnarray}\vspace{-4mm}
 \begin{eqnarray}
 \label{Eqn_Silent2_vx}
\end{eqnarray}}
\end{thm}
{\bf Proof:} 
Let us assume that agent $i$ succeeds at $\tau^i_1 \approx u$, i.e., say $\z_2 = (s, u)$.  It will then try to acquire  the second lock. In this case the optimal control problem would be (with   $x(u) := \ba^i (u)$, $a(u) = a_2^i (i)$   and $\Gamma := U-u$ and $T:= U$)

\vspace{-3mm}
{\small \begin{eqnarray*}
J(a (.) ) &=&  \hspace{-2mm} \int_0^{\Gamma}     c  \exp(-x (u) ) a (u) du\\&&\hspace{-7mm}   - \nu  \left (  \int_0^{\Gamma}   x(u)  \exp(-x(u) ) a (u) du  + x (\Gamma) \exp(-x (\Gamma) )   \right ).  
\end{eqnarray*}} 
This is exactly as in Theorem \ref{Thm_silent} corresponding to $M=1$ case and hence   $a^* \equiv  0$ (if $c < \nu$) or $a^* \equiv \bnd^i$  (if $c  \ge  \nu$) depending upon the value of   $\nu$. Note  the optimal control  does not depend upon $u$ or $\z_2$. However the value function depends upon $u$ and hence $\z_2.$

If $\nu > c$, because of the above result,  it is easy to observe that $a^* = \Gamma_2 (0)$, i.e., $a_1^{i*} =  0$ as well as $a_2^{i*} (\z_2) \equiv 0$, i.e.,  for all $0 \le t \le T.$
Now we consider the case with $\nu \le  c.$

Further the
 value function at $\z_2$,  by applying Theorem \ref{Thm_silent} with $x = 0$ (note that $x(t)$ represents the fictitious state for the optimal control problem solving the second stage of the dynamic programming equation  and hence starts with $x(0) = 0$)
  is given by:
\begin{align*}
v_2^i (\z_2; \Gamma_2(0))  = \left(c-\nu \right)\left( 1-\exp(-\bnd^i (T- u) )\right). 
\end{align*} 
 By hypothesis the reward upon success is $c $. 
Thus  from DP equations (\ref{Eqn_DPs}),  the optimal control for the start ($a_1^{i*} (\cdot)$)  is obtained by optimizing the following objective function: 
{\small\beq
J (a(.))\hspace{-3mm}&=& \hspace{-4mm} \int_0^T   \left(c-\nu \right)\left( 1-\exp(-\bnd^i(T-u))\right)\exp (- x (u) )  a (u) du  \\&&\hspace{-8mm} -\nu  \left (  \int_0^{T}   x(u)  \exp(-x(u) ) a (u) du  + x (T) \exp(-x (T) )   \right ),
\eeq} over all $a (\cdot) \in {\cal C}$.  This is again an optimal control problem. 
We drop the superscript $i$ in this proof, for simpler notations. Thus we need the
solution of the following (Hamiltonian Jacobi) HJB PDE, with $v(t,x)$ representing
the value function and with $v_t$, $v_x$ its partial derivatives

\vspace{-3mm}
{\footnotesize
\beq
 \max_{a \in [0, \bnd]} \Bigg \{  \Bigg [ \underbrace{  \left ( ( c -
\nu) - \nu \x \right ) \exp{(-\x)}-  (c - \nu)\exp\left (-x+\bnd(t-T)\right ) 
}_{L(x,t,a)/a} \Bigg ]\a \\   \hspace{-3mm} +
 \a v_x (t, \x) \Bigg \} +  v_t  (t, \x) = 0 \eeq}with boundary condition $v(T, x) = -\nu \x \exp (-\x)$.

\ignore
{\color{red} Wrong .... \\
{\bf Claim:}  We claim the following is a solution satisfying the above boundary
valued problem (when $\nu <c $): 
\beq
W(t, \x) &=& (c-\nu)  x \exp (-x +\bnd (t-T)   ) - \nu x \exp (-x)  +  ( ( c - \nu )
- \nu ) \exp (-x)  + \kappa \exp(-x + \bnd t)   \mbox{ with} \\
\kappa &=&  -(c-\nu)  x \exp (-\bnd T ) -  ( (c - \nu) - \nu  ) \exp (-\bnd T).
\eeq
Note that its partial derivatives are:
\beq
W_t (t, x) &=&   \left \{  \begin{array}{lll}
 (c-\nu)  x \exp (-x +\bnd (t-T)   )+ \bnd \kappa \exp(-x + \bnd t)  &\mbox{if }  t
\le t_1 \\
 -\nu x    \exp (-x)   \mbox{if }  t > t_1.
\end{array}  \right .  ,\\
W_x (t, x) &=&   (c-\nu)   \exp (-x +\bnd (t-T) )  (1 -   x )  -   ( (c-\nu) -\nu x)
   \exp (-x)  - \kappa \exp(-x + \bnd t),\\
 L(t, \x, \a)  + \a W_x (t, \x)  \  = \   ((c-\nu)  - \nu )  \exp (-x+\bnd (t-T)).
\hspace{-50mm}
\eeq}
{\bf Claim:}  We claim the following is a solution satisfying the above boundary
value problem, prove that the optimal control is a threshold policy $\Gamma(t_1)$
and further compute $t_1$: 

\vspace{-4mm}
{\scriptsize\beq
W(t, \x) &=&  \left \{  \begin{array}{lll}  ( c-\nu) \bnd t   \exp(-x + \bnd t -
\bnd T)  \\ + (c - 2 \nu    - \nu x ) \exp (-x )  + \kappa \exp(-x + \bnd t)   
&\mbox{if }  t \le t_1 \\
 -\nu x    \exp (-x)   & \mbox{if }  t > t_1,
\end{array}  \right .  \\ \mbox{ with} \\
\kappa &=&   -    ( c-\nu) \bnd t_1  \exp( - \bnd T)    -    (c - 2\nu)  \exp (-\bnd
t_1).
\eeq}
Note that its partial derivatives are:
{\scriptsize\beq
W_t (t, x) &=& \hspace{-2mm}\left \{  \begin{array}{lll}    ( c-\nu) \bnd (t  \bnd + 1)   \exp(-x
+ \bnd t - \bnd T) \\ \hspace{10mm}+    \bnd \kappa \exp(-x + \bnd t)   &\mbox{if } 
t \le t_1 \\
 0  & \mbox{if }  t > t_1,
\end{array}  \right .   \\
W_x (t, x) &=&\hspace{-2mm}\left \{  \begin{array}{lll}   - ( c-\nu) \bnd t   \exp(-x + \bnd t -
\bnd T)    \\   -   ( c-\nu  -\nu x)    \exp (-x)  - \kappa \exp(-x +
\bnd t)  &\mbox{if }  t \le t_1 \\
 (\nu x - \nu) \exp (-x)   & \mbox{if }  t > t_1.
\end{array}  \right . \eeq}
For $t\le t_1$ we have for any $a$

{\small\begin{eqnarray}
\nonumber
&&\hspace{-9mm}\frac{ L(t, \x, \a)  + \a W_x (t, \x) }{a}\\ &  = &   - ( c-\nu) ( \bnd t + 1)   
\exp(-x + \bnd t - \bnd T)  -   \kappa  \exp (-x+\bnd t ) \nonumber \\
 &=&   -  ( c-\nu) ( \bnd t + 1 -  \bnd t_1 )\exp(-x + \bnd t - \bnd T ) \nonumber \\&&  +  (c - 2
\nu) \exp(-x + \bnd t - \bnd t_1 ) \nonumber  \\
 &=&    \exp(-x + \bnd t - \bnd t_1 ) \nonumber \\&&
   \bigg (    (c - \nu) (\bnd (t_1 - t) - 1) 
\exp (-\bnd (T -t_1) )   +  (c - 2 \nu)   \bigg ).  \hspace{6mm}
\label{Eqn_second_term}
\end{eqnarray}}
Thus  if $c > 2\nu$ and if we set $t_1$ as below
\beq
\exp (-\bnd (T -t_1) )    = \frac{ (c - 2\nu)}{(c - \nu) } \mbox { or if } 
t_1 := T +  \frac{1}{\bnd} \log  \left ( \frac{c - 2\nu }{c - \nu}  \right ),
\eeq then\footnote{
The required term is positive for all $t \le t_1$ if and only if the second term (\ref{Eqn_second_term}) is positive for all such $t$, this term is decreasing in $t$ and hence would be positive for all such $t$, if it is positive/non-negative at $t= t_1$
}  $ L(t, \x, \a)  + \a W_x (t, \x)  > 0$ for all $t \le t_1$.
Further  for all $t > t_1$, for the same choice of $t_1$ 

\vspace{-3mm}    
{\small\beq
&&\hspace{-12mm}\frac{ L(t, \x, \a)  + \a W_x (t, \x) }{a}\\   &=&   ( c - 2 \nu)  \exp{(-\x)}-  (c -
\nu)\exp\left (-x+\bnd(t-T)\right )   \\
&< & \hspace{-1mm}\exp (-x)  \bigg ( (c- 2\nu)   - ( c- \nu)  \exp(\bnd(t_1-T)) \bigg )      \\
&=&  0. 
\eeq }
Thus 
$W(., .)$ satisfies the HJB PDE for all $t \le T$ with $a^*(t)  = \Gamma(t_1) =
\bnd^i 1_{t \le t_1} $, 
\beq
W_t (t, \x)   + L(t, \x, \a^*(t) )  + \a^*(t) W_x (t, \x) &=& 0 \eeq
 and also satisfies the boundary condition 
 \beq
W (T, \x)
+ \kappa \exp(-x + \bnd T)  
&= & - \nu \x \exp(-\x)   \mbox{ for any } x.
 \eeq
 Thus the so defined   $\a^* (t) $   and $\x^* (t) = \bnd^i  \min \{t,  t_1\}$ 
satisfy equation (5.7) of \cite[Theorem 5.1]{Feller}, when $\nu \le c/2$ 
 and hence forms the optimal control. 
 
 When $\nu > c/2$, using similar logic one can show that  $W(t,x) = -\nu x \exp(-x)$
(for all $t$) is the solution of the HJB PDE and  $\a^*(t)  = 0$ for all $t$. Rest
of the things follow from  the  solutions,  $W(\cdot, \cdot).$
\eop

\TR{}{ 
\onecolumn
}

{\bf Proof of Theorem \ref{Thm_NE_two_lock}:}
We   compute the best response against threshold policies of type $\Gamma_2 (\psi^j)$ for those $\psi^j$ which can potentially become a part of the Nash Equilibrium. 
To begin with we obtain the best response against silent opponent in Theorem \ref{Thm_Silence_twolock}.
 Theorem \ref{Thm_Silence_twolock} is  used to  solve  the HJB equations (that represent the best response against  $\Gamma_2 (\psi^j)$) with $c = \exp (-\bnd^j \psi^j)$ and for all  $t\geq {\psi^j} $. 
By Dynamic programming principle (as applied to optimal control problem), one can use the results (the value function)  of Theorem \ref{Thm_Silence_twolock}  as the boundary condition (i.e., as $v(\psi^j, x)$) to solve the problem for $t\leq {\psi^j} $.
\TR{The rest of the proof is available in \cite{TR}.  \eop

\ignore{
@@@@@@@@
 
\vspace{-4mm}
{\small
 \begin{eqnarray*}
\exp (-\bnd^j \psi^{i*} )  = \min   \left \{1,   \exp (-\bnd^i (T-\psi^{i*}) -\bnd^j  \psi^{i*} )   +  \frac{ \nu}{ \left(1-\nu \right)}  \right \}.
 \end{eqnarray*}}
and  $\psi^{j*}$ satisfy the following equation 

\vspace{-4mm}
{\small \beq
  \exp (-\bnd^i \psi^{j*} )  =    \min   \left \{1,  \exp (-\bnd^j (T-\psi^{j*}) -\bnd^i \psi^{j*})   + \frac{ \nu }{1-\nu} \right \}.
\eeq}

With interior solutions, the required assumptions are always satisfied.

W.l.g. say $\beta^j > \beta^i $ and further if,

 \vspace{-4mm}
{\small \beq
    \exp (-\bnd^j T )     + \frac{ \nu }{1-\nu}    \ge 1,
\eeq}
then there is no interior  solution and $\psi^{j*} = 0 $.   So interior point is possible only when the above condition is negated.

Since $\beta^i < \beta^j$,  we also have 
{\small \beq
    \exp (-\bnd^i T )     + \frac{ \nu }{1-\nu}    \ge 1,
\eeq}
But if even at $T$ we have the above

 \vspace{-4mm}
{\small \beq
    \exp (-\bnd^j T )     + \frac{ \nu }{1-\nu}    >   \exp(-\bnd^j T),
\eeq}
then there is no interior  solution and $\psi^{i*} = 0 $.   So interior point is possible only when the above condition is negated.

\eop}}{

Thus for  $t\leq {\psi^j} $  we need to solve the following  HJB PDE (for all $  t\leq {\psi^j}  $)
\begin{eqnarray*}
v_t +   \max_{a \in [0, \bnd] } \bigg  \{ \left(1-\nu \right)\exp (-\bnd^j t)\exp(-x) -  (1-\nu )\exp (-\bnd^i T)\exp (-\bnd^j t)\exp (\bnd^i  t  )\exp(-x) \\ -\nu x\exp(-x) + v_x \bigg  \} { \a}   &=  0    \end{eqnarray*}
with boundary condition  as given by Theorem \ref{Thm_Silence_twolock} (with $c :=\exp (-\bnd^j \psi^j )$, $\Gamma := T- \psi^j$):
\begin{eqnarray*}
v(\psi^j, x) &=& \left \{ \begin{array}{lllll}
( c - 2\nu ) \exp(-x) (1 -  \exp (  -\beta^i \Gamma ) ) \\
 -\left (  \nu x  + ( c-\nu) \bnd^i \theta_{\nu_2} \exp( - \bnd^i  \Gamma )  \right )   \exp(-x)    &\mbox{if }  \theta_{\nu_2}   > 0 \\
- \nu x \exp (-x)   & \mbox{else, } 
\end{array} \right .    \\
 \theta_{\nu_2}(\psi^j) & :=& \max \left \{ 0,   \Gamma +  \frac{1}{\bnd^i} \log  \left ( \frac{c - 2\nu }{c - \nu}  \right )  \right \} 1_{ c > 2 \nu}
\end{eqnarray*}
We would proceed only with those $\psi^j$ for which   $\theta_{\nu_2} (\psi^j) $ = 0.  That is,   we have 
\begin{eqnarray} 
\label{Eqn_to_satisfy}
 \exp(- \bnd (T-  \psi^j) )  \ge    \frac{c - 2\nu }{c - \nu}  .   
\end{eqnarray} Note that $c < 2 \nu$ condition is also captured by the above.
In either case, the terminal condition is  $- \nu x \exp (-x) .$
 We drop subscript $i$ for simpler notations
and  conjecture the following to be the solution (for the best response against $\Gamma_2(\psi^j)$ and with $t \le \psi^j$) and obtain appropriate $t_1$:
\begin{align}
\label{solution}
W(t,x)= \left \{ \begin{array}{llll}
-(1-\nu) \frac{\bnd}{\beta^j} \exp (-\beta T)\exp(-\beta^jt)\exp (\beta t)\exp(-x)-(\nu x + \nu)\exp(-x) \nonumber\\ \hspace{5cm}+ \frac{\beta}{\beta + \beta^j}(1-\nu)\exp(-\beta^jt)\exp(-x)   + \kappa_1\exp(-x + \bnd t)  &\mbox{if }   t \le t_1 \\
-\nu x \exp (-x )   &\mbox{else. }
\end{array} \right .
\end{align}
with
\beq
\kappa_1 &=&   - \frac{\bnd}{\bnd + \bnd^j } (1-\nu)\exp(-\bnd^j t_1 ) \exp(-\bnd t_1 )  \\
&&+(1-\nu)\frac{\bnd   }{\bnd^j} \exp(-\bnd^j t_1 )\exp(-\bnd T)   + \nu\exp(-\bnd t_1 ).
\eeq
The partial derivatives are:
\beq
W_t(t,x) &=&   \left \{ \begin{array}{llll}  -(1-\nu) \frac{\bnd (\bnd -\bnd^j)}{\beta^j} \exp (-\beta T)\exp(-\beta^jt)\exp (\beta t)\exp(-x) \nonumber\\ 
 -  \frac{\beta \bnd^j }{\beta + \beta^j}(1-\nu)\exp(-\beta^jt)\exp(-x)   + \bnd  \kappa_1\exp(-x + \bnd t) 
 &\mbox{if }   t \le t_1 \\
0
 &\mbox{else. }
\end{array} \right .
\\
W_x(t,x) &=&  \left \{ \begin{array}{llll}  (1-\nu) \frac{\bnd}{\beta^j} \exp (-\beta T)\exp(-\beta^jt)\exp (\beta t) \exp(-x) \\
  +\nu x\exp(-x) - \frac{\beta}{\beta + \beta^j} (1-\nu)\exp(-\beta^j t)\exp(-x) - \kappa_1\exp(-x + \beta t)
 &\mbox{if }   t \le t_1 \\   \\
(\nu x  - \nu  ) \exp (-x)
 &\mbox{else. }
\end{array} \right .\eeq
Thus for any $t \le t_1$ we have
\beq
\frac{L (t, x,  a) + a W_x (t, x)}{a}  &=&  (1-\nu) \frac{\bnd - \beta^j}{\beta^j} \exp (-\beta T)\exp(-\beta^jt)\exp (\beta t) \exp(-x) \\
&& + \frac{\beta^j}{\beta + \beta^j} (1-\nu)\exp(-\beta^j t)\exp(-x)   - \kappa_1\exp(-x + \beta t) \\
%
%
&&\hspace{-20mm} =  (1-\nu) \bigg  (  \frac{\bnd - \beta^j}{\beta^j} \exp(-\beta^jt) \exp (-\beta T)    \\  
&&+ \frac{\beta^j}{\beta + \beta^j} \exp(-\beta^j t)   \exp(-\bnd t )  + \frac{\bnd}{\bnd + \bnd^j }  \exp(-\bnd^j t_1 ) \exp(-\bnd t_1 ) \\
&& -    \frac{\bnd   }{\bnd^j} \exp(-\bnd^j t_1 ) \exp (-\beta T)  - \frac{ \nu }{1-\nu} \exp (- \bnd t_1)  \bigg  )  \exp(-x+\bnd t ) \\
\eeq
We need to choose a $t_1 > 0$ such that the following is positive for all $t \le t_1$
\begin{eqnarray}
\label{Eqn_first}
  - \nu \exp (- \bnd t_1)  - (1-\nu)   \exp(-\beta^jt) \exp (-\beta T) \\ \nonumber  + (1-\nu) \frac{\beta^j}{\beta + \beta^j} \exp(-\beta^j t)   \exp(-\bnd t ) \\  + (1-\nu) \frac{\bnd}{\bnd + \bnd^j }  \exp(-\bnd^j t_1 ) \exp(-\bnd t_1 )  \ge 0, \nonumber
\end{eqnarray}
and such that for all $t \ge t_1$ (by computing  $L (t, x,  a) + a W_x (t, x)$ with $t \ge t_1$)
\begin{eqnarray}
\label{Eqn_condition_for_secon_leg}
 \left(1-\nu \right)\exp (-\bnd^j t) -  (1-\nu )\exp (-\bnd^i T)\exp (-\bnd^j t)\exp (\bnd^i  t  )  - \nu \le 0.
\end{eqnarray}

The RHS of  (\ref{Eqn_first})  is  decreasing  in $t$ (the derivative of RHS\footnote{the derivative equals
$ (1-\nu)   \exp(-\beta^jt ) \beta^j  \bigg (  \exp (-\bnd^i T)   - \exp (-\bnd^i t)  \bigg ) < 0$
}  of (\ref{Eqn_first})  is negative) and thus we will effectively require that  (\ref{Eqn_first})  is satisfied at an appropriate $t_1$, i.e., we will require a $t_1$ such that
\begin{eqnarray}
\label{Eqn_first_t1}
  - \nu   - (1-\nu)   \exp(-\beta^j t_1) \exp (-\beta T)  \exp (\bnd t_1)    +  (1-\nu) \exp(-\beta^j t_1)     \ge 0 
\end{eqnarray}

The RHS of second equation  (\ref{Eqn_condition_for_secon_leg}) is also decreasing (the derivative\footnote{derivative  =
$\left(1-\nu \right)\exp (-\bnd^j t)  \left ( -\bnd^j + (\bnd^j - \bnd_i) \exp (-\bnd^i (T-t) ) \right ) =
\left(1-\nu \right)\exp (-\bnd^j t)  \bigg (  - \bnd^j ( 1 - \exp (-\bnd^i (T-t) ))   - \bnd_i \exp (-\bnd^i (T-t) )  \bigg  ) < 0
 $
} is  again negative) in $t$, so once again it suffices to check (\ref{Eqn_condition_for_secon_leg})  at   $t= t_1$.   So, we would have an interior $t_1 = \psi^{i*} $ ,  if $t_1$ exactly solves the following:
 \begin{eqnarray}
 \left(1-\nu \right)\exp (-\bnd^j t_1 )  =  (1-\nu )\exp (-\bnd^i T)\exp (-\bnd^j t_1)\exp (\bnd^i  t_1  )   + \nu .
 \label{Eqn_theta_nu_2} 
 \end{eqnarray}
The RHS of (\ref{Eqn_first_t1}) with $t_1 = T$ equals $-\nu$ which is never positive, thus $t_1 < T$.  
$$\boxed{\mbox{In summary either $\psi^{i*} = t_1 > 0$ exactly satisfies   (\ref{Eqn_theta_nu_2})
or it would be $\psi^{i*} =0$ if $1 - \exp (-\bnd^i T) \le \frac{\nu}{1-\nu}$,}}$$ 
from (\ref{Eqn_condition_for_secon_leg}), i.e.,   the best response would be to be silent throughout.  {\it Thus  the thresholds given in the hypothesis  of  the theorem  for $(\psi^{i*}, \psi^{j*})$ are the correct thresholds.}  

Thus if there exists a $\psi^j$ such that $\theta_{\nu_2} (\psi^j) = 0$, the best response of agent $i$ is given by  $\Gamma_2 (\psi^{i*})$ where 
$\psi^{i*}$ satisfies (\ref{Eqn_theta_nu_2}) or is 0.  Note that this best response is the same for all $\psi^j$ such that $\theta_{\nu_2} (\psi^j) = 0$.   To find a potential NE, we will need best response of agent $j$ against $\Gamma_2 (\psi^{i*})$.   But  one can apply the analysis obtained so far with  roles of agent $j$ and agent $i$    reversed.   
Thus  if  $t_1 := \psi^{i*}$ satisfies  (\ref{Eqn_to_satisfy})  (with $\beta^i$ and $\beta^j$ interchanged)  then a $t_2 = \psi^{j*}$ that satisfies   
the following  (see (\ref{Eqn_theta_nu_2}))
\beq
 \left(1-\nu \right)\exp (-\bnd^i t_2 )  =  (1-\nu )\exp (-\bnd^j T)\exp (-\bnd^i t_2)\exp (\bnd^j t_2  )   + \nu ,
\eeq
(or $\psi^{j*}=0$) is the best response against $\Gamma_2 (\psi^{i*})$. We are done if we show that such a $\psi^{j*}$ satisfies (\ref{Eqn_to_satisfy}).  But this is ensured by (\ref{Eqn_conditions_for_NE}) of hypothesis. 

We are left to work with boundary conditions, which are considered below case by case.

\underline{\bf Case 1:  When $\bm {  \exp (-\beta^j T) <  (1-2 \nu) / (1-\nu) }$  and $\bm {\exp (-\beta^i T) \ge   (1-2\nu) / (1-\nu) }$}
In this case we claim that $\psi^{i*} = 0$ and $\psi^{j*} = \theta_{\nu_2}^j$.   
From Silence Theorem \ref{Thm_Silence_twolock}, the best response
$$
BR_j (\Gamma_2 (0) ) =  \Gamma_2 (\theta_{\nu_2}^j ).
$$  Remains to show the other best response is 0, i.e., to one should find  $BR_i (\Gamma_2 (\theta_{\nu_2}^j ))$.   With $\psi^{j} = \theta_{\nu_2}^j$, the RHS of equation (\ref{Eqn_to_satisfy}) 
equals
\beq
\exp (-\beta^i (T- \psi^j) ) \ge \exp (-\beta^i T)  \ge    \frac{1 - 2 \nu}{1- \nu},
\eeq
thus  condition (\ref{Eqn_to_satisfy})  is satisfied (in fact this condition would be satisfied for any $\psi^j$). 
Again for the given case, (\ref{Eqn_condition_for_secon_leg}) is satisfied for all $t \ge t_1$ with $t_1 = 0$ and thus we have
$$BR_i (\Gamma_2 (\theta_{\nu_2}^j )) =  \Gamma_2 ( 0 ).$$
Further from Silence Theorem \ref{Thm_Silence_twolock} we have  
$$\exp (- \beta^j \psi^{j*} ) =  \exp (- \beta^j T ) \frac{1- 2\nu}{1 - \nu} , 
$$and  we also have  $\psi^{j*}  > 0$  ( as $ \exp (- \beta^j \psi^{j*} ) \le   [(1-2 \nu) / (1-\nu) ]^2 < 1$ ).

\underline{\bf Case 2: When $\bm { \exp (-\beta^j T) \ge  (1-2\nu) / (1-\nu) }$  and $\bm { \exp (-\beta^i T) \ge (1-2\nu) / (1-\nu) }$}  Using exactly similar logic as in Case 1, one can show that 
$\psi^{i*} = 0= \psi^{j*}.$
%
%
%
  \eop
}
\end{document}